# Superlens as matching device

## V.G.Veselago

Moscow Institute of Physics and Technology, Moscow district, Russia
Institute of General Physics, Moscow,Russia

The appearance of new class of material - material with negative refraction, which show many unusual electrodynamic properties, has brought intensive searching for their new characteristics and possible practical applications. Herewith some statements appear in literature, which cause motivated objections. So, in work [1] becomes firmly established that negative refraction exists for phase velocity only, but group velocity has usual law of refraction with positive value of refraction index $n$. The authors of this work are not embarrassed by the fact, that difference in directions of phase and group velocity is typical properties of optical anisotropic media, which can not be charaterized by scalar value of $n$. The mistake of authors [1] can be explained by the fact, that authors muddle the direction of group velocity with direction of perpendicular to surfaces of constant amplitude for modulated waves. This mistake is in detail considered and explained in paper [2].

There is one more problem, which is in close relation with appearance of materials with negative refraction. This is the problem of "overcoming of diffraction limit", or, but on several other terminology, problem of increasing of so-called evanescent modes. For the first time this problem was discussed by J.B.Pendry in work [3], where was shown that in material with negative refraction can successfully spread waves, for which component $k_z$ of wave vector along direction of spreading has imaginary value

$$k_z = i\sqrt{\frac{\omega^2}{c^2} - k_x^2} \qquad (1)$$

This inequality is valid for very large $k_x$, that is to say for very short waves. In material with positive value of $n$ the amplitude of such waves (the evanescent modes) in accordance with (1) will exponentially decreased along $z$ axis, and exactly this fact explains impossibility of image by optical systems of objects, with sizes noticeably smaller than wavelength. However in work [3] and in many followed articles authors suggest that in material with negative refraction index waves with large values of $k_x$ do not decrease, but increase. Hereunder becomes suggested, that it is possible to transfer the images with sizes much less than wavelength from one point of space to another. This suggestion was motivated by the possibility of resonance increasing of surface modes in material with negative refraction.

The author [3] enter the notion "superlense" for device, like shown on Fig.1, confirming that for this sort of device the classical restriction on diffraction limit is not valid. The authors of work [3], and many others, more late, drew a veil over fact that overcoming of diffraction limit automatically meant breach of uncertainty principle.
For our case uncertainty equation can be written as follows

$$k_x d \geq 2\pi \qquad (2)$$

Here $k_x$ is component of wave vector orthogonal to $z$ axis, along which spreads the wave, but $d$ there is transverse size of focused spot of light. Value $k_x$ can not be more, than wave vector $k_0$ in free space:

$$k_x < k_0 = \frac{\omega}{c} = \frac{2\pi}{\lambda} \qquad (3)$$

From (2) and (3) immediately follows:

$$d \geq \lambda \qquad (4)$$

The possibility of breach of diffraction limit is equivalent to statement about unacceptability of equation (4), or, more exactly, to unnecessary its execution. Such statement about unnecessary of its execution is exceedingly strong statement, much more strong than all another possible statements about any other characteristics of material with negative refraction.

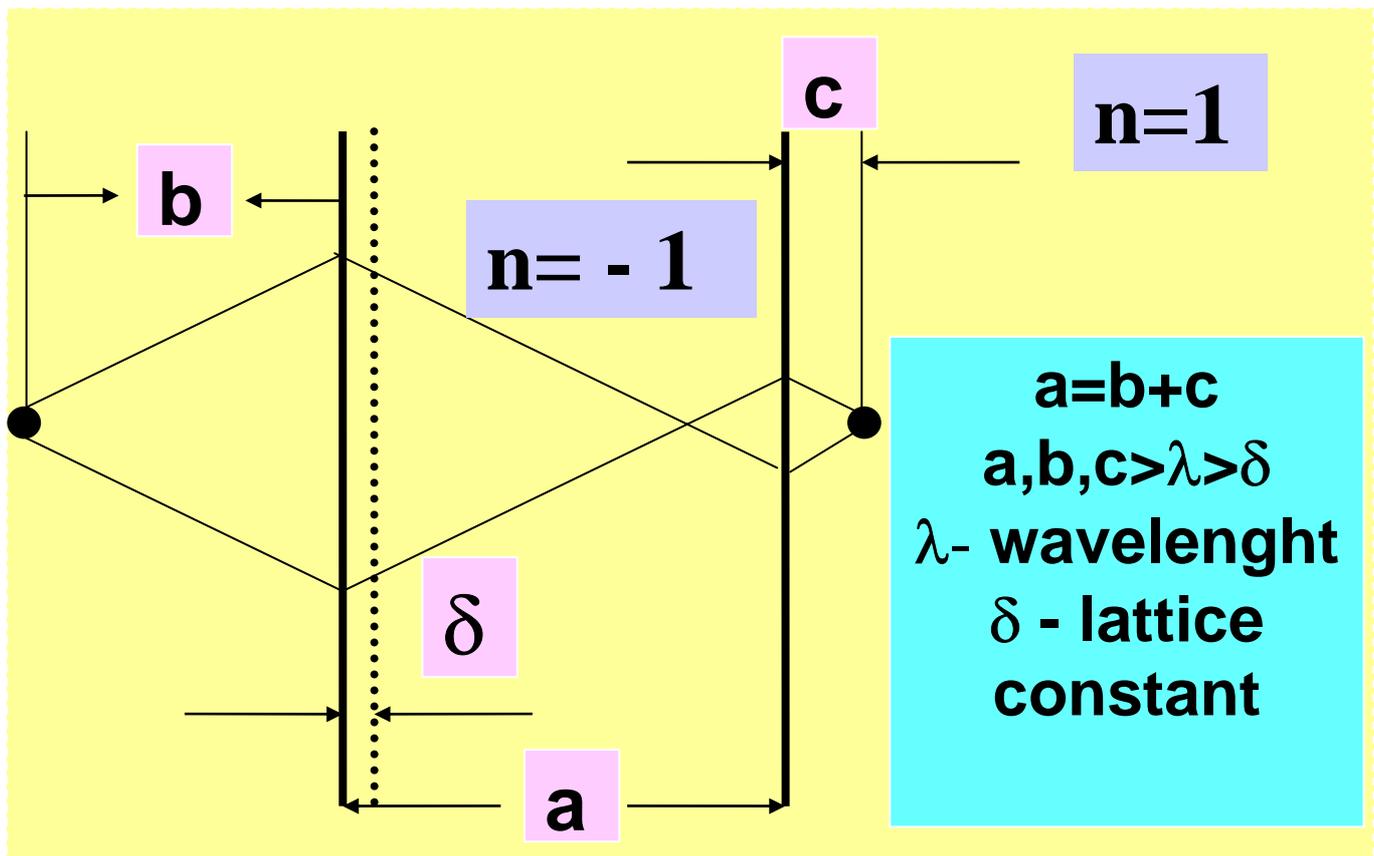

Fig. 1. The flat lens produced by material with refractions index $n = -1$.

On our opinion, fact of appearance of this sort of statement itself is from not very exact use of some terms, first of all main term - word "lens". This word "lens" itself characterizes the optical instrument, shown Fig.1, which work is founded on geometric optics laws.

However flat plate from material with negative refraction as is shown on Fig.1 can be considered as a lens, only if its transverse size $a$, wavelength of radiation $\lambda$ and period of internal structure $\delta$ satisfy to inequality

$$a > \lambda > \delta \qquad (5)$$

Besides, correlation must be executed in such lens

$$a = b + c \qquad (6)$$

Only in this case lens really works as optical instrument, complying with geometric optics laws. Exactly this situation was meant in our article [4] though this was not straight indicated.

However in basic article [3], and in the following works the situation was considered, when correlation (5) was not executed, since transverse size of lens there was one order with wavelength, and such order there was distances from lens to object and image planes. Such sort of a system is not a lens, but some matching device, which not at all will work on the base of geometric optics laws. As is well known, it is possible by means of matching devices to concentrate the flows of energy in to spot, undoubtedly smaller, than wavelength.

To make clear this situation, let us consider spreading and registration of electromagnetic wave in usual metallic waveguide. It is well known, that electromagnetic wave could spread in hollow rectangular waveguide with size $a$ of broad wall, if following relation is valid

$$\lambda \leq 2a \qquad (7)$$

If this condition is satisfied, wave spreads on waveguide with small fading, and field on output of such waveguide could be considered as some rectangle image, which is forming by cross-section of waveguide. The sizes of this rectangle are, on order of value, size $a$.

For registration of radiation, spreading in waveguide, one usually used the detectors, with sizes greatly less, than transverse size of waveguide. If such detector is situated beside output end of waveguide, it will register only small part of radiation, mainstream which will go by detector.

If, wanting enlarge the power, falling on detector, we will begin to narrow waveguide, that wave in waveguide will begin strongly to fade as soon as will is broken the correlation (7). However wholly possible greatly enlarge the power on detector, if use the different sort of matching devices, having placed them in waveguide in close proximity of detector. Such matching elements usually are a different sort of screws and slots, by means of which possible much powerfully enlarge the power, getting on detector. This increase power on detector is possible to consider as focusing of radiation to spot, which transverse size is a size of detector, and, naturally, it is greatly less than wavelength. Hereunder possible confirm that in waveguide with matching device is undoubtedly broken diffraction limit, and wave is focused on spot with size which noticeably less than wavelength. This really so, but only therefore that on detector fall already not only flat waves, which spread in the long waveguide. Except these flat waves, on detector acts the evanescent waves, which are generated by the elements of matching devices. These elements have the size less than wavelength, and placed on small distances from detector. Generation of evanescent waves itself results from reradiation of flat waves in waveguide on small details of matching devices.

In principle, just the same process exists in superlense, which is shown on Fig.1. This lens on essences of deal will match the radiator with receiver. Herewith are not executed the correlations (5) and (6), however lens indeed can send without garbling the scene with small sizes, but, regrettably, on small distances, comparable with wavelength [5].